\documentclass[useAMS,usenatbib]{mn2e}
\usepackage{ctable}
\usepackage{float}
\usepackage{graphicx}
\usepackage{color}
\usepackage{amsmath}
\usepackage{ctable}
\usepackage{caption}
\usepackage{subcaption}

\newcommand{\kms}{\ensuremath{\mathrm{km\,s^{-1}}}}

\newcommand{\sigmastar}{\sigma_{\star}}
\newcommand{\Mcutoff}{$M_{\mathrm{low} }$} 
\newcommand{\Msun}{M_\odot}

\newcommand{\MoLs}{\Upsilon_{\star}}

\newcommand{\Reff}{$R_{\rm eff }$}

\title[The low-mass IMF in ETGs]{Are the total mass density and the low-mass end slope of the IMF anti-correlated?\thanks{Based on observations collected 
at the European Organisation for Astronomical Research in the Southern Hemisphere, Chile. (P086.A-0312, PI: Koopmans  and P089.A-0364, PI: Spiniello)}}

\author[C.~Spiniello et al.]{C. Spiniello$^{1}$\thanks{E-mail:spini@mpa-garching.mpg.de}, 
M. Barnab{\`e}$^{2,3}$, L.V.E. Koopmans$^{4}$, S.C. Trager$^{4}$ 
\\
$^{1}$Max-Planck Institute for Astrophysics, Karl-Schwarzschild-Strasse 1, 85740 Garching, Germany\\
$^{2}$Dark Cosmology Centre, Niels Bohr Institute, University of Copenhagen, Juliane Maries Vej 30, 2100 Copenhagen {\O}, Denmark \\
$^{3}$Niels Bohr International Academy, Niels Bohr Institute, University of Copenhagen, Blegdamsvej 17, 2100 Copenhagen {\O}, Denmark \\
$^{4}$Kapteyn Astronomical Institute, University of Groningen, P.O. Box 800, 9700 AV Groningen, the Netherlands\\}
\begin{document}

%%%%%%%%%%%%%%%%%%%%%%%%%%%%%%%%%%%%%%%%%%%%%%%%
%\date{Accepted Year Month Day. Received Year Month Day; in original form Year Month Day}
\pagerange{\pageref{firstpage}--\pageref{lastpage}} \pubyear{2015}
\maketitle
\label{firstpage}

\begin{abstract}
We conduct a detailed lensing, dynamics and stellar population analysis of  
nine massive lens early-type galaxies (ETGs) from the X-Shooter Lens Survey 
(XLENS). Combining gravitational lensing constraints from \emph{HST} 
imaging with spatially-resolved kinematics and line-indices constraints 
from  \emph{VLT} X-Shooter (XSH) spectra, we infer the low-mass slope and 
the low cut-off mass of the stellar Initial Mass Function (IMF):
 $x_{250}=2.37^{+0.12}_{-0.12}$ and $M_{\mathrm{low}, 250}= 0.131^{+0.023}_{-0.026}\,\Msun$, 
respectively, for a reference point  with $\sigmastar \equiv 250\,\kms$ and \Reff $\equiv 10$ kpc. 
All the XLENS systems are consistent with an IMF slope steeper than 
Milky Way-like. 
We find no significant correlations between IMF slope and any other quantity,  except for an 
{\it anti-correlation} between total dynamical mass density and low-mass IMF slope at the 87\% CL
[$dx/d\log(\rho)$ = $ -0.19^{+0.15}_{-0.15}$]. %  or $dx/d\log(\rho) < 0 $ at the 87\% CL]. 
This anti-correlation is  consistent with the low redshift lenses found by Smith et al. (2015) 
that have high velocity dispersions and high stellar mass densities but surprisingly shallow IMF slopes.
\end{abstract}


\section{Introduction}
Low mass stars ($M \leq 0.3\, \Msun$) contribute at most 
5--10 per cent to the total optical light of the integrated spectrum of 
early-type galaxies (ETGs)  but dominate the total stellar 
mass budget (\citealt{Worthey1994}).  Variations in the number 
of M-dwarfs can therefore dramatically change the stellar 
mass-to-light ratio ($\MoLs$)  without significantly affecting 
the characteristics  of the observed light (e.g., \citealt{Bell2001, Conroy2012, Conroy2013}). 
Hence the initial mass function (IMF), and especially its low-mass slope, 
plays a fundamental role in determining an ETG's $\MoLs$ and is crucial 
to understand the internal structure, the formation times and the evolution of 
these massive galaxies (e.g.\ \citealt{Blumenthal1984}), 
providing important insight into galaxy 
evolution mechanisms (e.g. \citealt{Padmanabhan2004, Tortora2009,
Auger2010, Graves2010, Barnabe2011}).  \\
\indent Recent observations, based on gravitational lensing, dynamics and/or 
simple stellar population (SSP) modeling of galaxy spectra, indicate that the 
number of low-mass stars in the central region of ETGs 
increases rapidly with stellar velocity dispersion 
(\citealt{Treu2010, vanDokkum2010, Spiniello2011, 
Spiniello2012, Cappellari2012, LaBarbera2013,Tortora2013, 
Spiniello2014, Martin2015}). 
Recently, however, Smith, Lucey \& Conroy (2015, hereafter S15) 
presented results on three very massive lens ETGs
with high velocity dispersions ($\sigma > 300$ \kms) and  $\alpha$-enhanced 
abundances whose $\MoLs$ are compatible with a Milky 
Way-like IMF slope (although a Salpeter IMF cannot be excluded). 
The potential  tension between different studies could indicate that stellar 
mass (or velocity dispersion) is not the main driver
behind IMF variations, but only a proxy for other, more physically motivated causes.\\
\ctable[
caption = Properties of the nine XLENS systems used in the analysis., 
star,
label = xlensintro_obsSLACS
]{llllllll}{
\tnote[ ]{{\sl 1)} Based on WFPC2 F606W observations. {\sl 2)} Luminosity-weighted stellar kinematics of the lens extracted from a rectangular aperture of $2\arcsec \times 1\farcs5$ centered on the galaxy's center. See Paper II. {\sl 3)} Best-fit IMF slope from spectroscopic, line-index based SSP analysis. For three systems, the spectra do not have S/N high enough to securely constrain the IMF. %More details and inference on the other stellar population parameters will be presented in Paper IV, in prep.
}
 }{
\hline
\hline
{\bf XLENS System}  &{\bf z$_{\rm lens}$} &{\bf  z$_{\rm source}$} &{\bf  \Reff (kpc) }& {\bf $R_{\rm Ein}$ (kpc) } & {\bf $L_{\rm V}$ ($10^{11}L_{\odot})^{1}$} & {\bf $ \sigma_{\rm XSH}$ (km/s)$^{2}$} & {\bf IMF slope}$^{3}$ \\
%\textbf{ } &\textbf{ }  & \textbf{ }  & \textbf{ ('') } & \textbf{(kpc)} & \textbf{ }\\
\hline
SDSSJ0037$-$0942	&	0.1955	&	0.6322		&	7.03	&		4.95		&		 $1.09\pm0.06$ 		&	$ 277 \pm 6$  	& $2.5\pm0.25$ \\ %Reff in arcsec 2.19 mv 16.90 
SDSSJ0044$+$0113	&	0.1196	&	0.1965		&	5.56	&		1.72		&		 $0.61\pm0.05$		& $260 \pm 8$ 		&  $-$ \\ %2.61  16.32
SDSSJ0216$-$0813 	&	0.3317	&	0.5235		&	12.6	&		5.53		&		 $1.82\pm0.05$		& $ 327\pm 19$		&$2.9\pm 0.35$\\ %2.67  18.36  
SDSSJ0912$+$0029 	&	0.1642	&	0.3239		&	10.8	&		4.58		&		 $1.47\pm0.05$		&$ 325 \pm 10$ 		& $2.6\pm0.30$\\ %3.87  16.56
SDSSJ0935$-$0003 	&	0.3475	&	0.4670		&	20.7	&		4.26		&		 $2.07\pm0.07$		&$ 380 \pm 22$ 		&$-$\\ %4.24  17.71
SDSSJ0936$+$0913 	&	0.1897	&	0.5880		&	6.61	&		3.45		&		 $0.83\pm0.05$		&$ 256 \pm 18 $ 	&$-$\\ %2.11  17.12 
SDSSJ0946$+$1006 	&	0.2219	&	0.6085		&	8.33	&		4.95		&		 $0.66\pm0.06$		& $ 300 \pm 22 $	&$2.1\pm0.15$\\ %2.35  17.78
SDSSJ1143$-$0144 	&	0.1060	&	0.4019		&	9.21	&		3.27		&		 $0.95\pm0.05$		&$287 \pm 18 $ 		&$2.4\pm0.18$\\ %4.80  15.83 
SDSSJ1627$-$0053 	&	0.2076	&	0.5241		&	6.66	&		4.18		&		 $0.79\pm0.05$		& $ 303 \pm 23$		&$2.3\pm0.26$\\ %1.98  16.91 	
%SDSSJ2343$-$0030 	 &	0.1810	&	0.4630		&	8.27	&		4.62		&		17.17 		&$ 298 \pm 21 $		&XXX \\ %2.74
\hline
\hline
}
\indent Strong gravitational lensing combined with stellar 
dynamics (GL+D) allows one to measure stellar masses of galaxies independently 
from assumptions about their stellar IMF or knowledge of their stellar population. 
Therefore, when combined with a measurement of the galaxy luminosity, 
one obtains a measure of the  value of $\MoLs$ which is a function of the IMF, 
allowing the latter to be constrained (e.g. \citealt{Treu2010, Spiniello2011, 
Barnabe2012, Barnabe2013}).  
Applications of GL+D have shown that the total $\MoLs$ of massive ETGs increases monotonically 
with the velocity dispersion of the galaxy (e.g. \citealt{Grillo2009, Auger2010, Treu2010,Barnabe2011, Dutton2012}). 
A steepening of the low-mass end of the stellar IMF 
with galaxy mass could therefore be largely responsible  for this change in $\MoLs$, 
although a change in the dark matter (DM) fraction could also play a role (\citealt{Auger2009}).  \\
\indent With the X-Shooter Lens Survey (XLENS: \citealt{Spiniello2011}; \citealt{Barnabe2013}, 
hereafter B13) we aim to probe variations of the low-mass end 
of the IMF and disentangle stellar and DM in the internal region of lens 
ETGs, via a joint lensing+dynamics+stellar population analysis.   \\
\indent Except for B13, all studies to date have assumed that the lower mass limit of 
the IMF is constant, since varying this parameter does not strongly impact the spectra 
nor change the line-index measurements for any assumed IMF slope.  
Stars with masses below $\sim0.15 \, \Msun$ are however 
critical to determine~$\MoLs$. %, giving an important contribution to the total mass budget of the system (\citealt{Worthey1994, Conroy2013}).  
The only possible way to break this degeneracy is by obtaining a measurement of the total stellar 
mass from an independent method, which, as in B13, is provided by a combined 
lensing and dynamics analysis.  \\
\indent In this Letter we study the low-mass end properties of the stellar IMF, inferring its 
 slope and its low-mass cut-off (\Mcutoff) in nine massive lens ETGs, quadrupling the sample 
presented in B13, which we subsequently augment by three recently discovered 
low-redshift lenses (S15). These two sets allow us, for the first time,  to study 
trends of the IMF parameters (slope and low-mass limit) with stellar velocity dispersion, 
effective radius and galaxy mass density. 

\section{Data Set and Analysis}
The XLENS sample consists of eleven massive lens ETGs with $\sigmastar = 250 - 400$ \kms , 
all with multi-band \emph{HST} and \emph{VLT} XSH data. 
Here we use nine out of eleven systems with the necessary 
total luminosity and high signal-to-noise (S/N) spectra. 
%all the necessary ancillary data to perform a proper GL+D analysis.
Characteristics of the systems used in this work are presented in Table~1.  \\
\indent For the GL+D analysis we make use of the high-resolution \emph{F814W} images of the
lens systems, originally selected from the Sloan Lens ACS Survey (SLACS, \citealt{Bolton2008}). 
%We note that only nine systems had all the necessary ancillary data to perform a proper GL+D analysis.
For the SSP modeling, UVB--VIS high signal-to-noise spectra  
%high enough to perform stellar population analysis (
($\mathrm{S/N}>50$) have been obtained during two runs between 2011 and 2013 in slit mode
(UVB: $\mathrm{R}=3300$ with $1\farcs6\times 11\farcs$ slit; VIS:
$\mathrm{R}=5400$, with $1\farcs5 \times 11\farcs$ slit). \\
\indent We refer to Paper II (Spiniello et al. 2015, submitted) 
for a more detailed presentation of the full survey sample and 
the spatially resolved kinematics analysis of the systems, 
to B13 for a detailed description of the precise methodology
and to Paper III (Barnab{\`e} et al., in prep.) for a more extensive discussion and results of 
the GL+D analysis of the full sample as well as detailed  
constraints on the DM halo fraction and density profiles. 

%%%%%%%%%%%%%%%%%%%%%%%%%%%%%%%%%%%%%%%%%%%%%%%%%%%%%%%%%%%
\begin{figure*}
\centering
\begin{subfigure}{.5\textwidth}
  \centering
  \includegraphics[width=.84\linewidth]{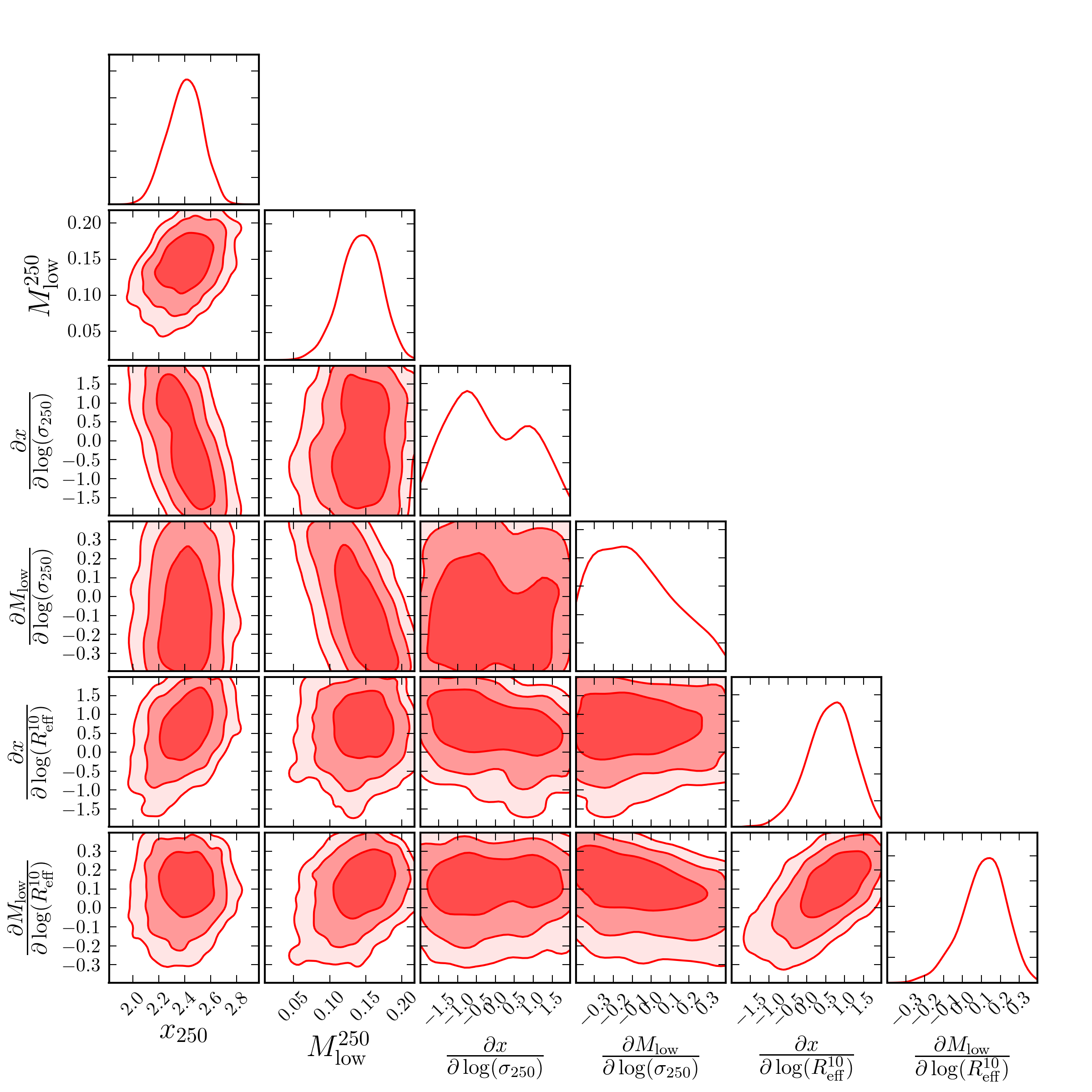}
\end{subfigure}%
\begin{subfigure}{.5\textwidth}
  \centering
  \includegraphics[width=.84\linewidth]{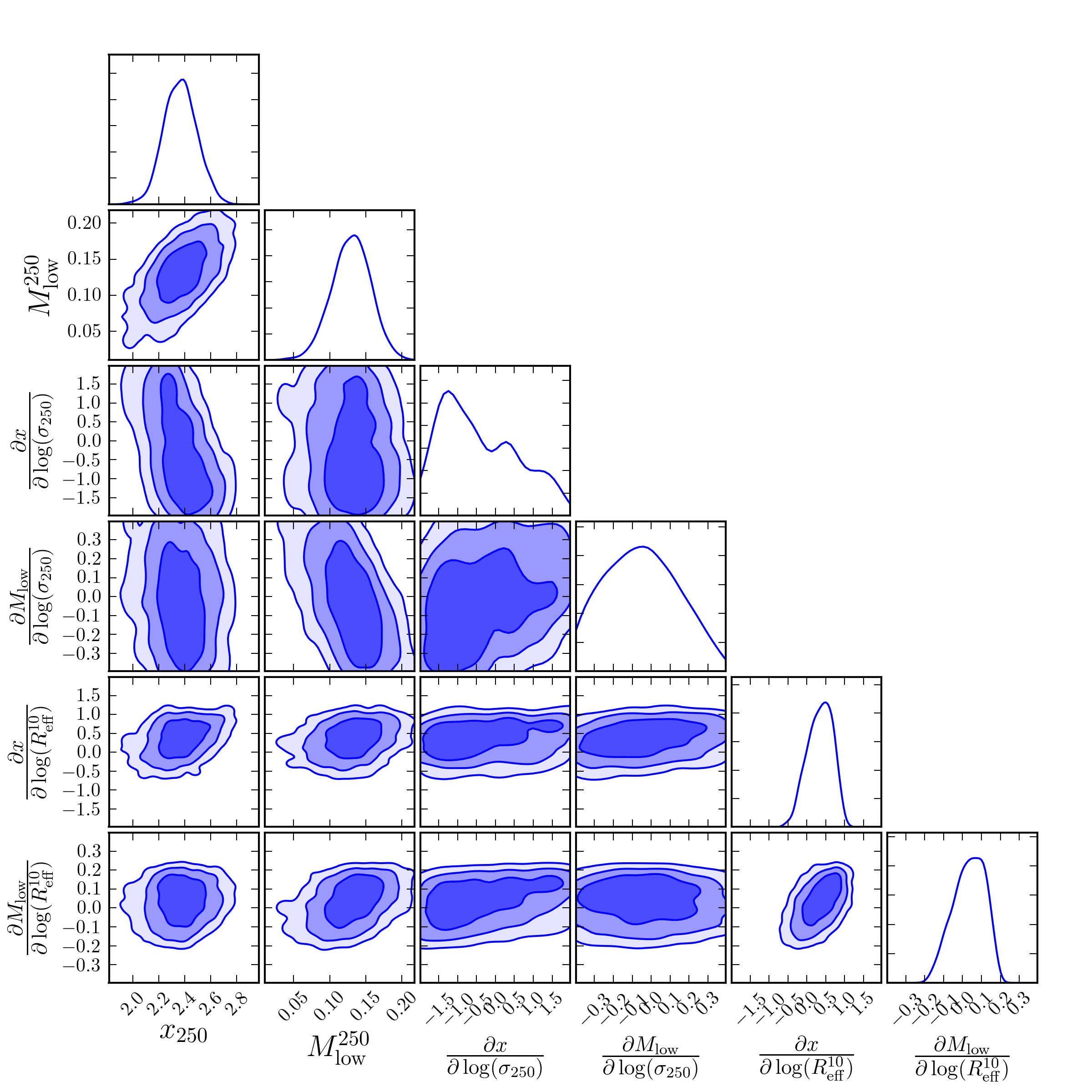}
\end{subfigure}
\caption{Marginalized posteriors for the model fitting of IMF slope ($x$) and cut-off mass 
(\Mcutoff) as function of $\sigma$ and $R_{\mathrm{eff}}$ (in unit of $250$ $\kms$ and 
10 kpc, respectively), including the XLENS systems only (left, red contours) 
or the XLENS$+$S15 systems (right, blue contours). }
\label{fig:mcmc_fitting}
\end{figure*}
%%%%%%%%%%%%%%%%%%%%%%%%%%%%%%%%%%%%%%%%%%%%%%%%%%%%%%%%%%%

%\section{Analysis}
To infer the IMF slope and cut-off mass, we apply the {same} methodology introduced in B13, 
summarized below. 

%\begin{itemize}
\indent i) From a GL+D analysis, we determine the posterior probability 
density function (PDF) of the stellar mass of each of the nine lens galaxies. 
This measurement is independent from any SSP or photometric analysis and 
does not make any assumption about the IMF. 
As in B13, we use the fully Bayesian
\textsc{CAULDRON} code (detailed in \citealt{Barnabe2007} and \citealt{Barnabe2012}) 
which is designed to conduct a self-consistent combined modelling of both the lensing 
and kinematic constraints. In particular we use a flexible two-component axially-symmetric 
mass model for each lens. We use a generalized Navarro-Frenk-White (gNFW) 
for the DM halo profile\footnote{The inner slope of the dark halo density
profile is allowed to vary between 0 and 2. This accounts for a wide
variety of possible DM inner slopes, from the pure NFW profile, to both 
steeper (`contracted') and shallower (`cored') profiles, proposed 
to try to include the effect of baryons on the dark halo inner regions.}, 
consistent with cosmological simulation findings, as well as observations, 
and we infer the density profile of the luminous mass component by deprojecting 
the multi-Gaussian expansion (MGE, \citealt{Emsellem1994, Cappellari2002}) fit to the observed surface brightness 
distribution of the galaxy. The total stellar mass then sets the normalization of the stellar mass distribution and is used to 
simultaneously model both the lensing data set and the stellar kinematic observables. \\ 
\indent ii)  We infer a value of the low-mass IMF slope from XSH spectra, 
using an extended version of the  SSP models of \citet{Conroy2012}, as described in 
\citet{Spiniello2014} and \citet{Spiniello2015}.   
This inference (see Table~1) is used as Gaussian prior in our subsequent analysis. 
Only six out of nine ETGs have spectra that provide a solid measurement. 
For the three remaining systems, 
we assume a flat prior on the IMF slope between $x=1.8$ and $3.5$.\\ 
\indent iii) We use a grid of $\MoLs$ values from the Dartmouth Stellar Evolution Program 
(\citealt{Chaboyer2001}) as functions of the IMF slope ($x=1.8 - 3.5$) and low-mass limit
(\Mcutoff$=0.01 - 0.22$). %, for each galaxy and its best SSP parameters. 
We use a Gaussian probability distribution function for total V-band luminosities ($L_{V}$) 
of the lens galaxies, as determined from \emph{HST} data by \citet{Auger2009}, to convert 
mass-to-light ratios ($\MoLs$) into stellar masses as function of the IMF 
slope and low-mass cut-off. \\ 
\indent iv) We subsequently use a Markov Chain Monte Carlo (MCMC) analysis of the model parameters
to derive the GL+D+SSP posterior PDFs of the IMF slope and low-mass cut-off  for each lens galaxy. 
In the model, $x$, $M_{\rm low}$ and $L_{V}$  all vary. The GL+D PDFs for the stellar mass produce the likelihood, 
which we then multiply with the priors on luminosity and IMF slope from the 
\emph{HST} and \emph{XSH} data, respectively. 
Finally, the luminosity of the galaxy is marginalized out. \\
\indent In this way, we obtain a posterior probability density for each system:
\begin{multline}
 P(x,M_{\rm low}|{\rm data}) = \int_{{\rm L}} {\cal{L}}({\rm data}_{{\rm GL+D}} | M_{*} =  \Upsilon_{V}(x,M_{\rm low}) \\ \times {\rm L}_{\rm V})  
\cdot P_{\rm HST}({\rm L}_{\rm V}) \cdot P_{\rm SSP}(x) \cdot P(M_{\rm low}) d{\rm L}_{\rm V}
\end{multline}
With these nine $P(x, M_{\rm low}| {\rm data}_{i})$ in hand, 
which are kept fixed in the subsequent analysis, we proceed with inferring whether there are 
any trends of the IMF parameters with stellar velocity dispersion ($\sigma$) and effective radius ($R_{\rm eff}$) 
or  the total (dynamical) density $\rho\equiv \sigma^2/R_{\rm eff}^2$. 
We first  construct a multivariate model: %described via the equations:
\begin{equation}
x   = x_{250} + x'_{1}\log_{10}(\sigma_{250}) + x'_{2} \log_{10}(R_{10})  
\end{equation}
\begin{equation}
M_{\mathrm{low}} = M_{\mathrm{low},250} + M_{\mathrm{low},1}^{' } \log_{10}(\sigma_{250}) + M_{\mathrm{low},2}^{' } \log_{10}(R_{10})  
\end{equation}
where $x_{250}$ and $M_{\mathrm{low}, 250}$ represent the  slope and the cut-off mass of the IMF respectively, 
and $\sigma_{250}$ and $R_{10}$ are the velocity dispersion in unit of $250$ $\kms$ and 
the effective radius in units of 10 kpc.  \\
\indent We then sample, using a MCMC Metropolis algorithm,  
all six free parameters (the two biases and four trends with velocity 
dispersion  and effective radius) and create their respective univariate and bivariate marginalised 
posteriors, assuming flat priors. The likelihood value is the product of the nine LHs 
values $\prod_{i=1, ... 9} P(x,M_{\mathrm{low}} | {\rm data}_{i})$ times the priors on the model 
parameters for the trend function inferred above (all assumed to be flat).
The small measurements errors on the effective radius, velocity dispersion and 
redshift are negligible in this process.  Each posterior is built from $10^5$ samples.  \\
%\indent Finally we marginalize the six-dimensional posterior to six one-dimensional PDFs, one for 
%each model parameter.  We also determine the two-variate (marginalized) posteriors which 
%allow degeneracies to be visualized.  The results are shown in the left panel of Figure~\ref{fig:mcmc_fitting}. \\
\indent Besides the nine XLENS galaxies, we augmented the MCMC 
analysis with three high-velocity dispersion, low-redshift lenses presented in S15.  
Using their constraints on the $\MoLs$ and the `mass excess factor' $\alpha$ (see Table 3 in S15),  
we infer the corresponding 
%(mass-to-light ratio mismatch relative to the Milky Way IMF), we extrapolate from the 
 value of the IMF slope (unimodal, single power-law) for each galaxy. 
%We assume a flat prior on the lower-mass limit to be conservative. 
%The right panel of Fig.~\ref{fig:mcmc_fitting} shows the results from the combined sample. 
This produces a tightening of most of the marginalised posteriors, 
but there is no major shift in the results, indicating that the model is 
consistent with both data-sets and their combination.   

\begin{figure*}
\centering
\begin{subfigure}{.5\textwidth}
  \centering
  \includegraphics[width=.84\linewidth]{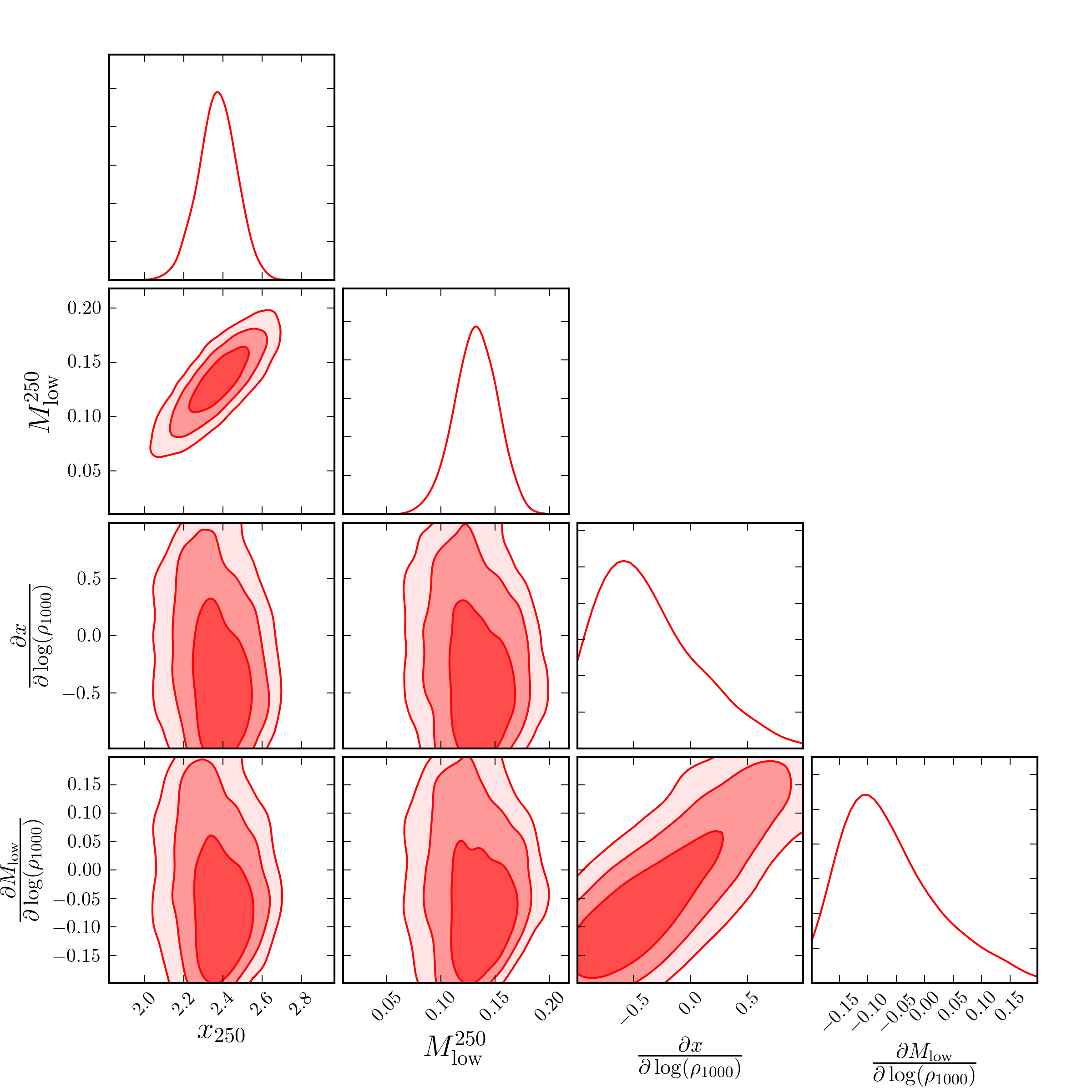}
\end{subfigure}%
\begin{subfigure}{.5\textwidth}
  \centering
  \includegraphics[width=.84\linewidth]{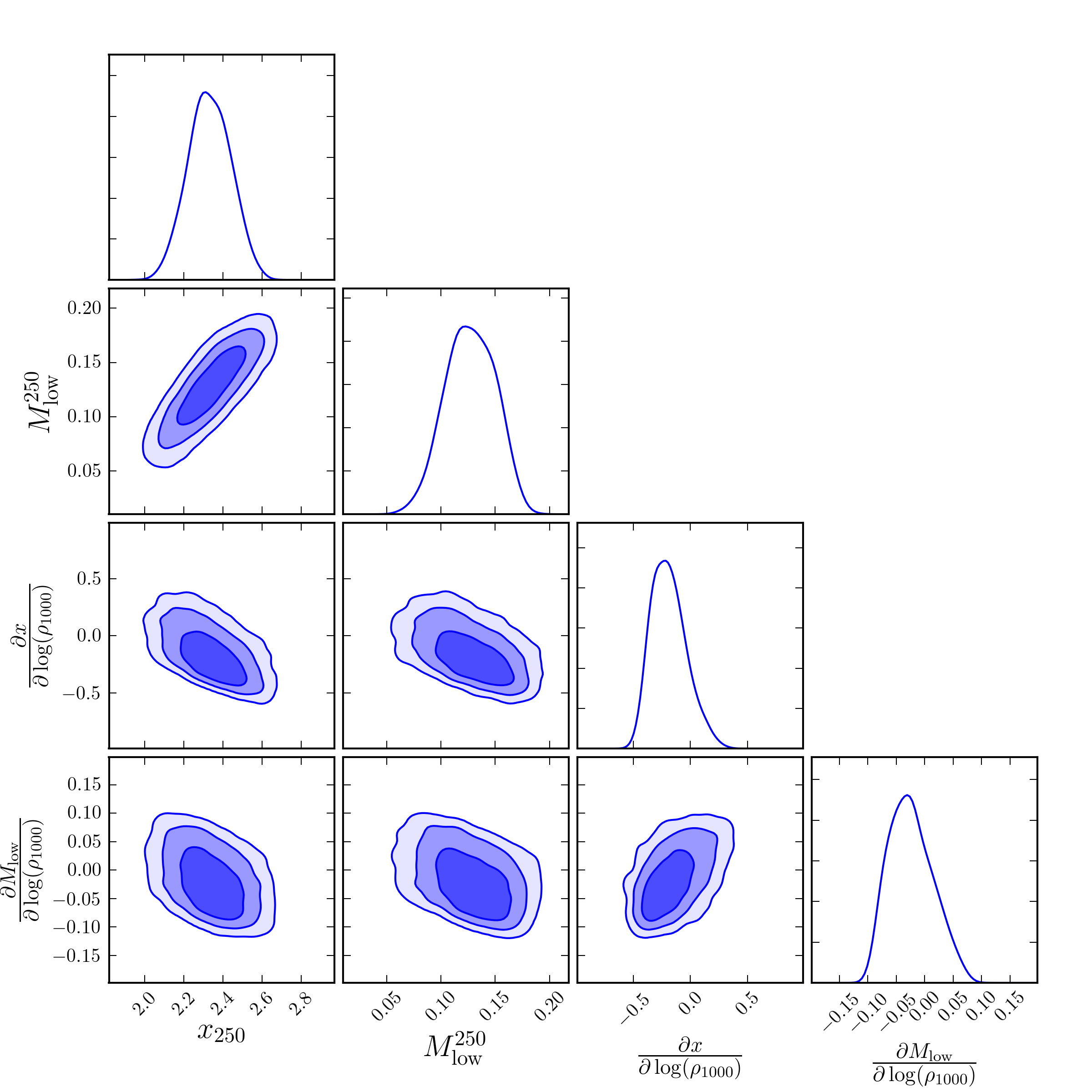}
\end{subfigure}
\caption{2-D marginalized PDFs for the model fitting of IMF slope ($x$) and low-mass cut-off (\Mcutoff) as function of total density 
($\rho=\sigma^{2}/R_{\mathrm{eff}}^2$) for the XLENS galaxies only (left, red contours) and for the XLENS+S15 sample (right, blue contours). 
The data suggest an anti-correlation between IMF slope and density.}
\label{fig:mcmc_rho}
\end{figure*}

\section{MCMC Results}
Figure~\ref{fig:mcmc_fitting} shows the univariate and bivariate marginalized posteriors resulting from the MCMC model fitting
of the nine XLENS (left, red plots) and of the XLENS+S15 systems (right, blue plots). 
The upper panels of the first two columns show the bias values of the slope and the low-mass cut-off 
defined for a reference system with $\sigma \equiv 250$ $\kms$ and $R_{\mathrm{eff}} \equiv 10$ kpc. %in the left panels of Figure~1.
We find $x_{250}$ = $2.41^{+0.11}_{-0.15}$ and $M_{\mathrm{low}, 250}$ = $0.144^{+0.024}_{-0.027}\Msun$ 
for XLENS and $x_{250}$ = $2.37^{+0.12}_{-0.12} $ and  
$M_{\mathrm{low}, 250}$ = $ 0.131^{+0.023}_{-0.026}\Msun$ for the XLENS+S15 case, respectively.
The results are fully consistent. In both cases a Milky Way-like IMF ($x=1.8$) is ruled out  
with a high degree of confidence  for our reference system, consistent with 
previous published results (B13, \citealt{Cappellari2012, Conroy2012b, Spiniello2014}). 
However, unlike previous results, we do not find any 
relation between the IMF slope and 
stellar velocity dispersion: $dx/d\log(\sigma)$ = $ -0.36^{+1.40}_{-0.89}$ for XLENS and 
$dx/d\log(\sigma)$ = $ -0.60^{+1.40}_{-0.89}$ for XLENS+S15.  
%The relation becomes even less significant 
%when we include the S15 systems: $\frac{\partial M_{\rm low}}{\partial \log(\sigma)}$ =$ -0.60^{+1.40}_{-0.89} $}. 
We note that the lack of a IMF-$\sigma$ relation
does not necessarily contradict previous studies, since this is the first time in which 
both the IMF slope and IMF low-mass cut-off  are treated as free parameters 
and the range of stellar velocity dispersions is relatively small ($280$-$380$ \kms).  
If we force a trend of $x$ with $\sigma$, we must also have a trend of 
$M_{\mathrm{low}}$ with $\sigma$. This indicates a degeneracy 
between the IMF slope and the IMF low mass cut-off, as expected. \\
\indent We find a marginally significant relation between  IMF slope and effective radius: 
 %bigger galaxies have steep IMF slopes at the 1-sigma 
$dx/d\log(R_{\rm eff})$ = $ 0.66^{+0.51}_{-0.59} $ for XLENS and 
$dx/d\log(R_{\rm eff})$ = $ 0.37^{+0.32}_{-0.36}$ for XLENS+S15. 
%\footnote{We  
%point out that we have ignored any correlation with redshift.}. 
%Conclusion from this is that $R_{\rm eff}$ might be a better indicator of the 
%IMF slope and cut-off mass, then the stellar velocity dispersion. 
We conclude, albeit with low significance, that the effective radius of ETGs might be a better indicator 
of IMF slope than their stellar velocity dispersion. Because in general the stellar dispersion correlates 
with effective radius, either can act as proxy for IMF slope, consistent with previous studies.

%\subsection{Total Mass Density}
Since both velocity dispersion and effective radius are global parameters, they might only be proxies 
for the physical processes during star-formation. On the other hand, mass 
density might be a better indicator of the IMF slope and the low-mass cut-off.  
Figure~\ref{fig:mcmc_rho} shows the marginalized posteriors resulting from a lower-dimensional 
MCMC power-law model fit where we define the total dynamical density as $\rho\propto \sigma^{2} / R_{\mathrm{eff}}^2$  
in units of $1000$ $(\kms)^{2}/\textrm {kpc}^{2}$. \\
\indent  We find that total density anti-correlates with the IMF slope, i.e. steeper 
slopes for less dense stellar systems, and that the  lower mass limit increases as well for 
less-dense systems: $dx/d\log(\rho)$ = $ -0.47^{+0.55}_{-0.33} $ 
for XLENS only and $dx/d\log(\rho)$ = $ -0.19^{+0.15}_{-0.15} $ 
for XLENS+S15 ($dx/d\log(\rho)< 0$ at 87\% CL). 
The relation is marginally significant, but when plotting the SSP-only inferred slopes 
of XLENS+S15 systems against total density  (Fig. 3), we see that the 
S15 lenses are all much denser and extend the anti-correlation trend already seen 
from the XLENS systems. Moreover,  $dx/d\log(\rho)$ obtained with 
a simple least-squares fit of the data (solid lines, red for XLENS only and blue for the XLENS+S15 case)
agrees well with that obtained from the MCMC analysis (dotted lines, same color code), 
but we emphasize that these fits are only a sanity-check of the MCMC results. 

\section{Summary \& Discussion}
\label{sec:summary}
We combine the results of a state-of-the-art lensing and dynamics 
analysis of nine massive lens galaxies of the XLENS Survey 
with corresponding inferences from a
spectroscopic SSP study of line-strength indices focusing on the 
properties of their IMFs. 
At the time of writing, this is the only method that allows the joint inference 
of the slope and low-mass cut-off of the IMF (see B13). \\
\indent We use a multi-variate, six-dimensional Markov Chain Monte Carlo analysis 
to derive the GL+D+SSP posterior PDFs on trends of the IMF slope and low-mass cut-off 
with stellar velocity dispersion, effective radius and total mass density. 
The main results of our analysis are as follows: \begin{enumerate}
\item For a reference galaxy with $\sigma \equiv 250$ $\kms$ and $R_{\mathrm{eff}}\equiv 10$ kpc, we infer 
$x_{250}$ = $2.37^{+0.12}_{-0.12} $ and $M_{\mathrm{low}, 250}$ = $0.131^{+0.023}_{-0.026}$ $M_{\odot}$, consistent
with the results previously published in B13 and with other published studies  on massive ETGs. 
%\item A MCMC analysis shows no correlation between the IMF properties and stellar velocity 
%dispersion, when both $x$ and \Mcutoff\, are allowed to vary. 
\item A marginally significant correlation is found with the size of the galaxies, 
i.e. systems with larger $R_{\mathrm{eff}}$ prefer bottom-heavier IMFs. 
This correlation is similar to that with stellar velocity dispersion because 
more massive galaxies, statistically, also have larger effective radii. 
\item We find an anti-correlation between IMF slope and total (dynamical) density, 
defined as $\rho=\sigma^{2} / R_{\mathrm{eff}}^2$. This trend could 
explain the seemingly contradictory result by \citet{Smith2015} that suggests that 
some very high-velocity dispersion lenses have shallow IMFs slopes, counter to previous findings. 
Combining the XLENS+S15 sample, we find that the IMF slope anti-correlates (at 87\% CL) 
with total mass density: $dx/d\log(\rho)$ = $-0.19^{+0.15}_{-0.15}$. 
\end{enumerate}
The fully Bayesian analysis of the IMF parameters 
using the XLENS systems confirms our previous results in B13 that the IMF slope is 
inconsistent with that of the Milky Way at a very high confidence level for reference 
galaxies with $\sigma \equiv 250$ \kms\, and \Reff $\equiv 10$ kpc. 
We find marginal ($\sim1$-$\sigma$) trends with effective radius and total mass-density. 
No trend with stellar velocity dispersion is found (although this is still consistent with our 
previous work).  When we include the sample of S15, both trends become stronger;  
in particular the anti-correlation of the total mass density with IMF slope is found at the 87\% CL. 
The S15 galaxies are much denser, with high stellar dispersion ($\sigma=[356\pm16, 356\pm 18, 320\pm 18]$) but smaller 
effective radii ($R_{\rm eff, J}(kpc) =[6.6,2.0,6.0]$), than the XLENS galaxies.  
%\footnote{This could possibly be a selection bias such that the low redshift 
%S15 lenses are easier to recognize, if their effective radii are small compared their Einstein 
%radii (providing a better contrast of source against lens.}. 
If indeed this trend with  
mass density is genuine, it explains the seemingly discrepant results between most studies and that by S15, 
but it also implies that a more physical (anti)correlation might exist between the low-mass IMF 
slope and the inner total mass density of ETGs. The reason for this anti-correlation remains 
an open question but it is likely related to the gas density during star formation, 
as suggested by recent theoretical papers (e.g.,  \citealt{Krumholz2011, Hopkins2012, Narayanan2013}).

\begin{figure} 
\center
\includegraphics[height=6.4cm]{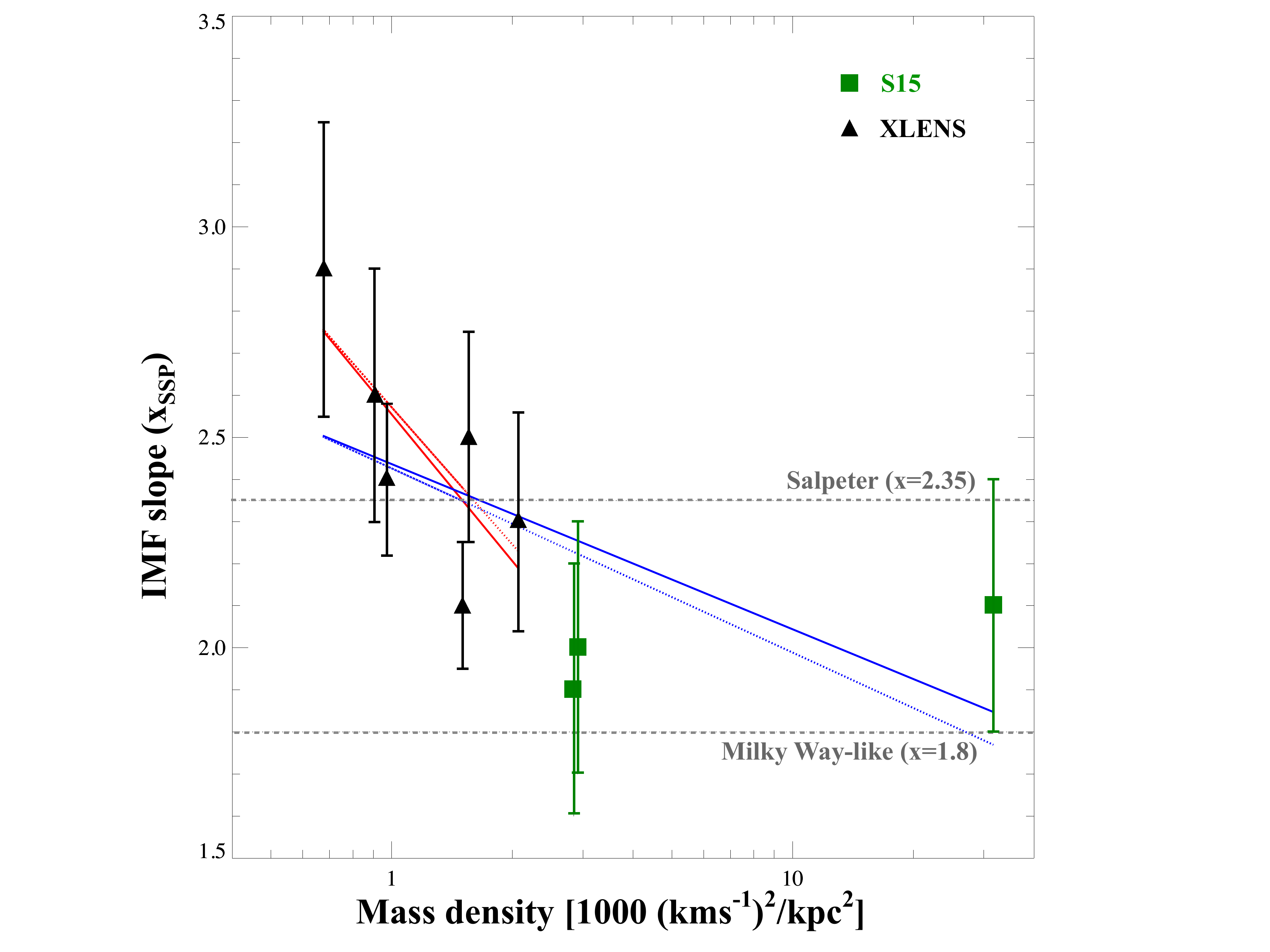}
\caption{IMF slopes 
of XLENS (black triangles)  and S15 systems (green squares) against total density. Solid lines show a least-square fit of the data, dotted lines show the 
relations obtained from the MCMC analysis. Red lines show the XLENS only case, blue lines show the XLENS+S15 case. The IMF slope plotted here is the one listed in Table.1, obtained via SSP analysis alone. }
\label{fig:IMF-dens}
\end{figure}

%The analysis performed in this work represents an improvement with respect
%to the pilot program of B13, since now we are also investigating possible trends of both the slope and, for the first time, 
%the low-mass end of the IMF with stellar velocity dispersion,  galaxy size, and total mass density. 

\section*{Acknowledgments}
We thank C.Grillo and S.Vegetti for thoughtful comments that have improved the quality of the manuscript. 
LVEK is supported in part through an NWO-VICI career grant (project number 639.043.308).

%%%%%%%%%%%%%%%%%%%%%%%%%%%%%%%%%%%%%%%%%%%%%%%%%%%%%%%%
% INPUT BIBLIOGRAPHY
%%%%%%%%%%%%%%%%%%%%%%%%%%%%%%%%%%%%%%%%%%%%%%%%%%%%%%%%
\bibliographystyle{mn2e_fix}
\bibliography{full_phd_Kiara.bib}

\label{lastpage}

\end{document}